\documentclass[aps,prl,twocolumn,showpacs,superscriptaddress,groupedaddress]{revtex4}  
\usepackage{graphicx}  
\usepackage{dcolumn}   
\usepackage{bm}        
\usepackage{amssymb}   
\usepackage{amsmath}
\usepackage{color}
\usepackage{multirow}
\usepackage{rotating}
\usepackage{xspace}
\usepackage{lscape}
\usepackage{comment}
\usepackage{slashed}

\newcommand{\fix}[1]{} 
\newcommand{\MET}{\mbox{$\raisebox{.3ex}{$\not\!$}E_T$}}
\newcommand{\METVEC}{\mbox{$\raisebox{.3ex}{$\not\!$}{\vec E}_T$}}

\newcommand{\ttbar}     {\mbox{$t\bar{t}$}\xspace}

\newcommand{\ljets}     {\mbox{$\ell$+jets}\xspace}
\newcommand{\comphep}   {\sc comphep}
\newcommand{\singletop} {\sc singletop}
\newcommand{\pythia}    {\mbox{\textsc{pythia}}}

\newcommand{\geant}     {{\sc{geant}}}
\newcommand{\alpgen}    {\mbox{\textsc{alpgen}}}
\newcommand{\mcfm}      {\sc mcfm}

\newcommand{\ifb}       {fb$^{-1}$}

\newcommand{\xsectev}{1.29}
\newcommand{\xsecteverrorup}{+0.26}
\newcommand{\xsecteverrordown}{-0.24}

\newcommand{\probtev}{1.8 \times 10^{-10}}
\newcommand{\sigmatev}{6.3}
\newcommand{\sigmaexp}{5.1}

\begin{document}

\hspace{5.2in} \mbox{FERMILAB-PUB-14-031-E}

\title{Observation of {\boldmath$s$}-channel production of single top quarks
  at the Tevatron}
\affiliation{Institute of Physics, Academia Sinica, Taipei, Taiwan 11529, Republic of China}
\affiliation{Argonne National Laboratory, Argonne, Illinois 60439, USA}
\affiliation{University of Athens, 157 71 Athens, Greece}
\affiliation{Institut de Fisica d'Altes Energies, ICREA, Universitat Autonoma de Barcelona, E-08193, Bellaterra (Barcelona), Spain}
\affiliation{Baylor University, Waco, Texas 76798, USA}
\affiliation{Istituto Nazionale di Fisica Nucleare Bologna, \ensuremath{^{uu}}University of Bologna, I-40127 Bologna, Italy}
\affiliation{University of California, Davis, Davis, California 95616, USA}
\affiliation{University of California, Los Angeles, Los Angeles, California 90024, USA}
\affiliation{Instituto de Fisica de Cantabria, CSIC-University of Cantabria, 39005 Santander, Spain}
\affiliation{Carnegie Mellon University, Pittsburgh, Pennsylvania 15213, USA}
\affiliation{Enrico Fermi Institute, University of Chicago, Chicago, Illinois 60637, USA}
\affiliation{Comenius University, 842 48 Bratislava, Slovakia; Institute of Experimental Physics, 040 01 Kosice, Slovakia}
\affiliation{Joint Institute for Nuclear Research, RU-141980 Dubna, Russia}
\affiliation{Duke University, Durham, North Carolina 27708, USA}
\affiliation{Fermi National Accelerator Laboratory, Batavia, Illinois 60510, USA}
\affiliation{University of Florida, Gainesville, Florida 32611, USA}
\affiliation{Laboratori Nazionali di Frascati, Istituto Nazionale di Fisica Nucleare, I-00044 Frascati, Italy}
\affiliation{University of Geneva, CH-1211 Geneva 4, Switzerland}
\affiliation{Glasgow University, Glasgow G12 8QQ, United Kingdom}
\affiliation{Harvard University, Cambridge, Massachusetts 02138, USA}
\affiliation{Division of High Energy Physics, Department of Physics, University of Helsinki, FIN-00014, Helsinki, Finland; Helsinki Institute of Physics, FIN-00014, Helsinki, Finland}
\affiliation{University of Illinois, Urbana, Illinois 61801, USA}
\affiliation{The Johns Hopkins University, Baltimore, Maryland 21218, USA}
\affiliation{Institut f\"{u}r Experimentelle Kernphysik, Karlsruhe Institute of Technology, D-76131 Karlsruhe, Germany}
\affiliation{Center for High Energy Physics: Kyungpook National University, Daegu 702-701, Korea; Seoul National University, Seoul 151-742, Korea; Sungkyunkwan University, Suwon 440-746, Korea; Korea Institute of Science and Technology Information, Daejeon 305-806, Korea; Chonnam National University, Gwangju 500-757, Korea; Chonbuk National University, Jeonju 561-756, Korea; Ewha Womans University, Seoul, 120-750, Korea}
\affiliation{Ernest Orlando Lawrence Berkeley National Laboratory, Berkeley, California 94720, USA}
\affiliation{University of Liverpool, Liverpool L69 7ZE, United Kingdom}
\affiliation{University College London, London WC1E 6BT, United Kingdom}
\affiliation{Centro de Investigaciones Energeticas Medioambientales y Tecnologicas, E-28040 Madrid, Spain}
\affiliation{Massachusetts Institute of Technology, Cambridge, Massachusetts 02139, USA}
\affiliation{University of Michigan, Ann Arbor, Michigan 48109, USA}
\affiliation{Michigan State University, East Lansing, Michigan 48824, USA}
\affiliation{Institution for Theoretical and Experimental Physics, ITEP, Moscow 117259, Russia}
\affiliation{University of New Mexico, Albuquerque, New Mexico 87131, USA}
\affiliation{The Ohio State University, Columbus, Ohio 43210, USA}
\affiliation{Okayama University, Okayama 700-8530, Japan}
\affiliation{Osaka City University, Osaka 558-8585, Japan}
\affiliation{University of Oxford, Oxford OX1 3RH, United Kingdom}
\affiliation{Istituto Nazionale di Fisica Nucleare, Sezione di Padova, \ensuremath{^{vv}}University of Padova, I-35131 Padova, Italy}
\affiliation{University of Pennsylvania, Philadelphia, Pennsylvania 19104, USA}
\affiliation{Istituto Nazionale di Fisica Nucleare Pisa, \ensuremath{^{ww}}University of Pisa, \ensuremath{^{xx}}University of Siena, \ensuremath{^{yy}}Scuola Normale Superiore, I-56127 Pisa, Italy, \ensuremath{^{zz}}INFN Pavia, I-27100 Pavia, Italy, \ensuremath{^{aaa}}University of Pavia, I-27100 Pavia, Italy}
\affiliation{University of Pittsburgh, Pittsburgh, Pennsylvania 15260, USA}
\affiliation{Purdue University, West Lafayette, Indiana 47907, USA}
\affiliation{University of Rochester, Rochester, New York 14627, USA}
\affiliation{The Rockefeller University, New York, New York 10065, USA}
\affiliation{Istituto Nazionale di Fisica Nucleare, Sezione di Roma 1, \ensuremath{^{bbb}}Sapienza Universit\`{a} di Roma, I-00185 Roma, Italy}
\affiliation{Mitchell Institute for Fundamental Physics and Astronomy, Texas A\&M University, College Station, Texas 77843, USA}
\affiliation{Istituto Nazionale di Fisica Nucleare Trieste, \ensuremath{^{ccc}}Gruppo Collegato di Udine, \ensuremath{^{ddd}}University of Udine, I-33100 Udine, Italy, \ensuremath{^{eee}}University of Trieste, I-34127 Trieste, Italy}
\affiliation{University of Tsukuba, Tsukuba, Ibaraki 305, Japan}
\affiliation{Tufts University, Medford, Massachusetts 02155, USA}
\affiliation{University of Virginia, Charlottesville, Virginia 22906, USA}
\affiliation{Waseda University, Tokyo 169, Japan}
\affiliation{Wayne State University, Detroit, Michigan 48201, USA}
\affiliation{University of Wisconsin, Madison, Wisconsin 53706, USA}
\affiliation{Yale University, New Haven, Connecticut 06520, USA}
\affiliation{LAFEX, Centro Brasileiro de Pesquisas F\'{i}sicas, Rio de Janeiro, Brazil}
\affiliation{Universidade do Estado do Rio de Janeiro, Rio de Janeiro, Brazil}
\affiliation{Universidade Federal do ABC, Santo Andr\'{e}, Brazil}
\affiliation{University of Science and Technology of China, Hefei, People's Republic of China}
\affiliation{Universidad de los Andes, Bogot\'{a}, Colombia}
\affiliation{Charles University, Faculty of Mathematics and Physics, Center for Particle Physics, Prague, Czech Republic}
\affiliation{Czech Technical University in Prague, Prague, Czech Republic}
\affiliation{Institute of Physics, Academy of Sciences of the Czech Republic, Prague, Czech Republic}
\affiliation{Universidad San Francisco de Quito, Quito, Ecuador}
\affiliation{LPC, Universit\'{e} Blaise Pascal, CNRS/IN2P3, Clermont, France}
\affiliation{LPSC, Universit\'{e} Joseph Fourier Grenoble 1, CNRS/IN2P3, Institut National Polytechnique de Grenoble, Grenoble, France}
\affiliation{CPPM, Aix-Marseille Universit\'{e}, CNRS/IN2P3, Marseille, France}
\affiliation{LAL, Universit\'{e} Paris-Sud, CNRS/IN2P3, Orsay, France}
\affiliation{LPNHE, Universit\'{e}s Paris VI and VII, CNRS/IN2P3, Paris, France}
\affiliation{CEA, Irfu, SPP, Saclay, France}
\affiliation{IPHC, Universit\'{e} de Strasbourg, CNRS/IN2P3, Strasbourg, France}
\affiliation{IPNL, Universit\'{e} Lyon 1, CNRS/IN2P3, Villeurbanne, France and Universit\'{e} de Lyon, Lyon, France}
\affiliation{III. Physikalisches Institut A, RWTH Aachen University, Aachen, Germany}
\affiliation{Physikalisches Institut, Universit\"{a}t Freiburg, Freiburg, Germany}
\affiliation{II. Physikalisches Institut, Georg-August-Universit\"{a}t G\"{o}ttingen, G\"{o}ttingen, Germany}
\affiliation{Institut f\"{u}r Physik, Universit\"{a}t Mainz, Mainz, Germany}
\affiliation{Ludwig-Maximilians-Universit\"{a}t M\"{u}nchen, M\"{u}nchen, Germany}
\affiliation{Panjab University, Chandigarh, India}
\affiliation{Delhi University, Delhi, India}
\affiliation{Tata Institute of Fundamental Research, Mumbai, India}
\affiliation{University College Dublin, Dublin, Ireland}
\affiliation{Korea Detector Laboratory, Korea University, Seoul, Korea}
\affiliation{CINVESTAV, Mexico City, Mexico}
\affiliation{Nikhef, Science Park, Amsterdam, the Netherlands}
\affiliation{Radboud University Nijmegen, Nijmegen, the Netherlands}
\affiliation{Joint Institute for Nuclear Research, RU-141980 Dubna, Russia}
\affiliation{Institution for Theoretical and Experimental Physics, ITEP, Moscow 117259, Russia}
\affiliation{Moscow State University, Moscow, Russia}
\affiliation{Institute for High Energy Physics, Protvino, Russia}
\affiliation{Petersburg Nuclear Physics Institute, St. Petersburg, Russia}
\affiliation{Instituci\'{o} Catalana de Recerca i Estudis Avan\c{c}ats (ICREA) and Institut de F\'{i}sica d'Altes Energies (IFAE), Barcelona, Spain}
\affiliation{Uppsala University, Uppsala, Sweden}
\affiliation{Taras Shevchenko National University of Kyiv, Kiev, Ukraine}
\affiliation{Lancaster University, Lancaster LA1 4YB, United Kingdom}
\affiliation{Imperial College London, London SW7 2AZ, United Kingdom}
\affiliation{The University of Manchester, Manchester M13 9PL, United Kingdom}
\affiliation{University of Arizona, Tucson, Arizona 85721, USA}
\affiliation{University of California Riverside, Riverside, California 92521, USA}
\affiliation{Florida State University, Tallahassee, Florida 32306, USA}
\affiliation{Fermi National Accelerator Laboratory, Batavia, Illinois 60510, USA}
\affiliation{University of Illinois at Chicago, Chicago, Illinois 60607, USA}
\affiliation{Northern Illinois University, DeKalb, Illinois 60115, USA}
\affiliation{Northwestern University, Evanston, Illinois 60208, USA}
\affiliation{Indiana University, Bloomington, Indiana 47405, USA}
\affiliation{Purdue University Calumet, Hammond, Indiana 46323, USA}
\affiliation{University of Notre Dame, Notre Dame, Indiana 46556, USA}
\affiliation{Iowa State University, Ames, Iowa 50011, USA}
\affiliation{University of Kansas, Lawrence, Kansas 66045, USA}
\affiliation{Louisiana Tech University, Ruston, Louisiana 71272, USA}
\affiliation{Northeastern University, Boston, Massachusetts 02115, USA}
\affiliation{University of Michigan, Ann Arbor, Michigan 48109, USA}
\affiliation{Michigan State University, East Lansing, Michigan 48824, USA}
\affiliation{University of Mississippi, University, Mississippi 38677, USA}
\affiliation{University of Nebraska, Lincoln, Nebraska 68588, USA}
\affiliation{Rutgers University, Piscataway, New Jersey 08855, USA}
\affiliation{Princeton University, Princeton, New Jersey 08544, USA}
\affiliation{State University of New York, Buffalo, New York 14260, USA}
\affiliation{University of Rochester, Rochester, New York 14627, USA}
\affiliation{State University of New York, Stony Brook, New York 11794, USA}
\affiliation{Brookhaven National Laboratory, Upton, New York 11973, USA}
\affiliation{Langston University, Langston, Oklahoma 73050, USA}
\affiliation{University of Oklahoma, Norman, Oklahoma 73019, USA}
\affiliation{Oklahoma State University, Stillwater, Oklahoma 74078, USA}
\affiliation{Brown University, Providence, Rhode Island 02912, USA}
\affiliation{University of Texas, Arlington, Texas 76019, USA}
\affiliation{Southern Methodist University, Dallas, Texas 75275, USA}
\affiliation{Rice University, Houston, Texas 77005, USA}
\affiliation{University of Virginia, Charlottesville, Virginia 22906, USA}
\affiliation{University of Washington, Seattle, Washington 98195, USA}

\author{T.~Aaltonen\ensuremath{^{\dag}}}
\affiliation{Division of High Energy Physics, Department of Physics, University of Helsinki, FIN-00014, Helsinki, Finland; Helsinki Institute of Physics, FIN-00014, Helsinki, Finland}
\author{V.M.~Abazov\ensuremath{^{\ddag}}}
\affiliation{Joint Institute for Nuclear Research, RU-141980 Dubna, Russia}
\author{B.~Abbott\ensuremath{^{\ddag}}}
\affiliation{University of Oklahoma, Norman, Oklahoma 73019, USA}
\author{B.S.~Acharya\ensuremath{^{\ddag}}}
\affiliation{Tata Institute of Fundamental Research, Mumbai, India}
\author{M.~Adams\ensuremath{^{\ddag}}}
\affiliation{University of Illinois at Chicago, Chicago, Illinois 60607, USA}
\author{T.~Adams\ensuremath{^{\ddag}}}
\affiliation{Florida State University, Tallahassee, Florida 32306, USA}
\author{J.P.~Agnew\ensuremath{^{\ddag}}}
\affiliation{The University of Manchester, Manchester M13 9PL, United Kingdom}
\author{G.D.~Alexeev\ensuremath{^{\ddag}}}
\affiliation{Joint Institute for Nuclear Research, RU-141980 Dubna, Russia}
\author{G.~Alkhazov\ensuremath{^{\ddag}}}
\affiliation{Petersburg Nuclear Physics Institute, St. Petersburg, Russia}
\author{A.~Alton\ensuremath{^{\ddag}}\ensuremath{^{ii}}}
\affiliation{University of Michigan, Ann Arbor, Michigan 48109, USA}
% ...

\collaboration{CDF Collaboration}
\altaffiliation[With visitors from]{
\ensuremath{^{a}}University of British Columbia, Vancouver, BC V6T 1Z1, Canada,
\ensuremath{^{b}}Istituto Nazionale di Fisica Nucleare, Sezione di Cagliari, 09042 Monserrato (Cagliari), Italy,
\ensuremath{^{c}}University of California Irvine, Irvine, CA 92697, USA,
\ensuremath{^{d}}Institute of Physics, Academy of Sciences of the Czech Republic, 182 21, Czech Republic,
\ensuremath{^{e}}CERN, CH-1211 Geneva, Switzerland,
\ensuremath{^{f}}Cornell University, Ithaca, NY 14853, USA,
\ensuremath{^{g}}University of Cyprus, Nicosia CY-1678, Cyprus,
\ensuremath{^{h}}Office of Science, U.S. Department of Energy, Washington, DC 20585, USA,
\ensuremath{^{i}}University College Dublin, Dublin 4, Ireland,
\ensuremath{^{j}}ETH, 8092 Z\"{u}rich, Switzerland,
\ensuremath{^{k}}University of Fukui, Fukui City, Fukui Prefecture, Japan 910-0017,
\ensuremath{^{l}}Universidad Iberoamericana, Lomas de Santa Fe, M\'{e}xico, C.P. 01219, Distrito Federal,
\ensuremath{^{m}}University of Iowa, Iowa City, IA 52242, USA,
\ensuremath{^{n}}Kinki University, Higashi-Osaka City, Japan 577-8502,
\ensuremath{^{o}}Kansas State University, Manhattan, KS 66506, USA,
\ensuremath{^{p}}Brookhaven National Laboratory, Upton, NY 11973, USA,
\ensuremath{^{q}}Queen Mary, University of London, London, E1 4NS, United Kingdom,
\ensuremath{^{r}}University of Melbourne, Victoria 3010, Australia,
\ensuremath{^{s}}Muons, Inc., Batavia, IL 60510, USA,
\ensuremath{^{t}}Nagasaki Institute of Applied Science, Nagasaki 851-0193, Japan,
\ensuremath{^{u}}National Research Nuclear University, Moscow 115409, Russia,
\ensuremath{^{v}}Northwestern University, Evanston, IL 60208, USA,
\ensuremath{^{w}}University of Notre Dame, Notre Dame, IN 46556, USA,
\ensuremath{^{x}}Universidad de Oviedo, E-33007 Oviedo, Spain,
\ensuremath{^{y}}CNRS-IN2P3, Paris, F-75205 France,
\ensuremath{^{z}}Universidad Tecnica Federico Santa Maria, 110v Valparaiso, Chile,
\ensuremath{^{aa}}The University of Jordan, Amman 11942, Jordan,
\ensuremath{^{bb}}Universite catholique de Louvain, 1348 Louvain-La-Neuve, Belgium,
\ensuremath{^{cc}}University of Z\"{u}rich, 8006 Z\"{u}rich, Switzerland,
\ensuremath{^{dd}}Massachusetts General Hospital, Boston, MA 02114 USA,
\ensuremath{^{ee}}Harvard Medical School, Boston, MA 02114 USA,
\ensuremath{^{ff}}Hampton University, Hampton, VA 23668, USA,
\ensuremath{^{gg}}Los Alamos National Laboratory, Los Alamos, NM 87544, USA,
\ensuremath{^{hh}}Universit\`{a} degli Studi di Napoli Federico I, I-80138 Napoli, Italy
}
\noaffiliation
\collaboration{D0 Collaboration}
\altaffiliation[With visitors from]{
\ensuremath{^{ii}}Augustana College, Sioux Falls, SD, USA,
\ensuremath{^{jj}}The University of Liverpool, Liverpool, UK,
\ensuremath{^{kk}}DESY, Hamburg, Germany,
\ensuremath{^{ll}}Universidad Michoacana de San Nicolas de Hidalgo, Morelia, Mexico,
\ensuremath{^{mm}}SLAC, Menlo Park, CA, USA,
\ensuremath{^{nn}}University College London, London, UK,
\ensuremath{^{oo}}Centro de Investigacion en Computacion - IPN, Mexico City, Mexico,
\ensuremath{^{pp}}Universidade Estadual Paulista, S\~{a}o Paulo, Brazil,
\ensuremath{^{qq}}Karlsruher Institut f\"{u}r Technologie (KIT) - Steinbuch Centre for Computing (SCC),
\ensuremath{^{rr}}Office of Science, U.S. Department of Energy, Washington, D.C. 20585, USA,
\ensuremath{^{ss}}American Association for the Advancement of Science, Washington, D.C. 20005, USA,
\ensuremath{^{tt}}National Academy of Science of Ukraine (NASU) - Kiev Institute for Nuclear Research (KINR)
}
\noaffiliation

\date{February 20, 2014}

\begin{abstract}

We report the first observation of single-top-quark production in the
$s$ channel through the combination of the CDF and D0 measurements of
the cross section in proton-antiproton collisions at a center-of-mass energy
of 1.96~TeV. The data correspond to total integrated luminosities of
up to 9.7~\ifb\ per experiment.  The measured cross section is 
$\sigma_s = \xsectev^{\xsecteverrorup}_{\xsecteverrordown}$~pb.
The probability of observing a statistical fluctuation of the
background to a cross section of the observed size or larger is
$\probtev$, corresponding to a significance of $\sigmatev$ standard
deviations for the presence of an $s$-channel contribution to the
production of single-top quarks.

\end{abstract}

\pacs{14.65.Ha; 12.15.Ji; 13.85.Qk; 12.15.Hh}
\maketitle



The top quark, with a mass of
$m_t=173.2\pm0.9$\;~GeV~\cite{topmassTeV}, is the most massive and one of
the most puzzling elementary particles of the standard model
(SM). Detailed studies of top-quark production and decay provide
powerful tests of strong and 
electroweak interactions, as well as sensitivity to physics beyond the
standard model (BSM)~\cite{theory}.  At the Tevatron, where
protons ($p$) and antiprotons ($\bar{p}$) collide at a center-of-mass energy of
$\sqrt{s}=1.96$\;TeV, top quarks are produced predominantly in pairs
(\ttbar) via the strong interaction~\cite{ttbar_combi}.  Top quarks
are also produced singly 
in $p{\bar p}$ collisions via the electroweak interaction. The
single-top-quark production cross section is expected to be
proportional to the square of the magnitude of the quark-mixing
Cabibbo-Kobayashi-Maskawa matrix~\cite{ckm} element $V_{tb}$, and
consequently sensitive to potential contributions from a fourth 
generation of quarks~\cite{FourthGen1,FourthGen2}, as well as
flavor-changing neutral currents~\cite{FCNC1,FCNC2,FCNC_CDF,FCNC_D0}, 
anomalous top-quark
couplings~\cite{anomcoup,anomcoup_D0_1,anomcoup_D0_2}, heavy $W'$ 
bosons~\cite{WPrime1,WPrime2,WPrime_CDF,WPrime_D0}, supersymmetric
charged Higgs   
bosons~\cite{ChargedHiggs,ChargedHiggs_D0}, or other new
phenomena~\cite{Tait:2000sh,DM_CDF}. 

\begin{figure}[!h!tbp]
\begin{center}
\includegraphics[width=0.48\textwidth]{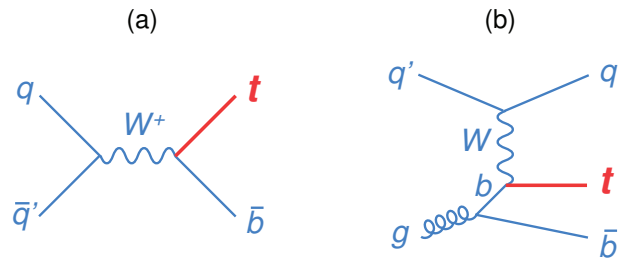}
\caption{Dominant Feynman diagrams for (a) $s$-channel
  and (b) $t$-channel 
single-top-quark production at the Tevatron.}
\label{fig:feynman_diagrams}
\end{center}
\end{figure}

At the Tevatron, there are two important processes in which a
single top quark is produced in association with other quarks. The
dominant channel proceeds through the exchange of a space-like virtual
$W$~boson between a light quark and a bottom quark ($b$ quark) in the
$t$~channel~\cite{singletop-willenbrock,singletop-yuan,tchannel-kidonakis}.
A second mode occurs through the exchange of a time-like virtual
$W$~boson in the $s$~channel, which produces a top quark and a
$b$ quark~\cite{singletop-cortese}.
Figure~\ref{fig:feynman_diagrams} shows the leading Feynman diagrams for the
$s$- and $t$-channel production modes.
%
%
Independent measurements of $s$-channel and $t$-channel production are
important, since BSM contributions could have different effects
on the two modes~\cite{Tait:2000sh}.

Single-top-quark production, independent of channel, was reported by
the CDF and D0 collaborations in
Refs.~\cite{stop-obs-2009-cdf,cdf-prd-2010}  
and~\cite{stop-obs-2009-d0,stop-2011-d0}, respectively. The D0
collaboration subsequently measured with larger data sets the
production cross section for the combined $s$ and $t$
channels~\cite{d0_schannel}, and  obtained $\sigma_{s+t} =
4.11^{+0.59}_{-0.55}$\;pb using a data set of 9.7~\ifb\ in agreement
with the SM prediction of $3.15 \pm 0.19$~pb
($m_t=172.5$~GeV)~\cite{schannel-kidonakis,tchannel-kidonakis}. 

After establishing 
the $s+t$ process, the cross sections of the individual production
modes were measured independently. 
Several differences in the properties of $s$- and $t$-channel events can be used
to distinguish them from one another.  Events originating from $t$-channel
production typically contain one light-flavor jet in the forward detector region
(at large pseudorapidity), which is useful for distinguishing them from events
associated with $s$-channel production and other SM background processes.
Moreover, events from the $s$-channel process are more likely to contain two
jets originating from $b$ quarks ($b$ jets) within the central region
of the detector where 
they can be identified.  Hence, single-top-like events with two
identified $b$ jets
are more likely to have originated from $s$-channel production.
Exploiting these differences, the D0
collaboration observed the $t$-channel process~\cite{t-channel-new},
and measured its cross section to be $\sigma_t = 2.90\pm 
0.59$\;pb. This compares to the SM prediction of $2.10
\pm 0.13$~pb ($m_t=172.5$~GeV)~\cite{tchannel-kidonakis}.
At the CERN LHC proton-proton collider, $t$-channel
production was also observed by the ATLAS and CMS
collaborations~\cite{atlas-tchannel,cms-tchannel}.  

Observing the $s$-channel process is more difficult, since the
expected cross section is smaller  than that of the $t$ channel and its
kinematic features are less distinct from the background. However,
the Tevatron has an advantage over the LHC in this mode, since valence
quarks ($q\bar{q}'$ from $p\bar{p}$) generally initiate $s$-channel
single-top-quark production, leading to a larger signal-to-background
ratio at the Tevatron than at the LHC. Due to this advantage, the CDF
and D0  collaborations have reported evidence for $s$-channel
production independently of each
other~\cite{cdf_schannel,d0_schannel}, while the LHC experiments have 
to date reported only unpublished upper limits on the cross section.


In this Letter, we report a combination of $s$-channel cross section
analyses performed by the CDF~\cite{cdf_schannel,cdf_schannel_MET} 
and D0~\cite{d0_schannel} collaborations. The CDF and D0 detectors are
central magnetic spectrometers surrounded by electromagnetic and
hadronic calorimeters and muon detectors~\cite{CDFII,D0II,note}.  The
combined measurement utilizes the full Tevatron Run II data sets
corresponding to up to $9.7$\;fb$\rm ^{-1}$ of integrated luminosity
per experiment. 

The data are selected using a logical OR of many
online selection requirements which preserve high signal efficiency
for offline analysis. 
Since the magnitude of the $W$-top-bottom quark coupling is much
larger than the $W$-top-down and $W$-top-strange quark
couplings~\cite{pdg}, each top quark decays almost exclusively to a
$W$~boson and a $b$ quark. The selection is split into two distinct final-state
topologies, both designed to select single-top-quark events in which the
$W$~boson decays leptonically.

One final-state topology (\ljets), analyzed by both
collaborations, contains 
single-top-quark events in which the $W$~boson decays leptonically ($W
\to \ell \nu_\ell$). We select events that (i)
contain only one isolated lepton ($\ell = e$ or $\mu$) with large transverse
momentum $p_T$, (ii) have large missing transverse energy
\MET~\cite{note}, (iii) have either two jets (CDF analysis) or  
two or three jets (D0 analysis) with large $p_T$, and (iv) have one or two $b$
jets. To identify $b$ jets, multivariate techniques are used that
discriminate $b$ jets from jets originating from 
light quarks and gluons~\cite{CDFbtag,D0btag}. Additional selection
criteria are applied to exclude kinematic regions that are
difficult to model, and to minimize the quantum chromodynamics (QCD)
multijet background where one jet is misreconstructed as a lepton and
spurious \MET\ arises from jet energy mismeasurements. 

The other final-state topology, analyzed by the CDF collaboration,
involves~\MET\ and jets, but no reconstructed 
isolated charged leptons (\MET+jets). The CDF analysis
avoids overlap with the \ljets sample by explicitly vetoing events
with identified leptons~\cite{cdf_schannel_MET}.  Large missing
transverse energy is required and events with two or three
reconstructed jets are accepted. This additional sample increases the
acceptance for $s$-channel signal 
events by encompassing those in which the $W$-boson decay produces a
muon or electron that is either not reconstructed or not isolated, or
a hadronically decaying
tau lepton that is reconstructed as a third jet.  After the basic
event selection, QCD multijet events dominate the \MET+jets event
sample. To reduce this multijet background, a neural-network
event selection is optimized to preferentially select signal-like
events.

Events passing the \ljets and \MET+jets selections are further
separated into independent analysis channels based on the number of
reconstructed jets as well as the number and quality of $b$-tagged
jets.  Each of the analyzed channels has a different
background composition and signal ($s$) to
background ($b$) ratio. Analyzing them separately enhances the
sensitivity to single-top-quark 
production~\cite{cdf_schannel,d0_schannel,cdf_schannel_MET}.


Both collaborations use Monte Carlo~(MC) generators to simulate the
kinematic properties of signal and background events, except in the
case of multijet production, for which the model is derived from data.
The CDF analysis models single-top-quark signal events at 
next-to-leading-order (NLO) accuracy in the strong coupling constant
$\alpha _s$ using the 
\textsc{powheg}~\cite{POWHEG2009} generator. 
The D0 analysis uses the {\singletop}~\cite{singletop-mcgen} event
generator, based on NLO 
{\comphep} calculations that 
match the event kinematic features predicted by NLO
calculations~\cite{singletop-xsec-sullivan,Campbell:2009ss}.
Spin information in the decays of the top quark and the $W$~boson is
preserved for both \textsc{powheg} and {\singletop}.

Kinematic properties of background events associated with the $W$+jets
and $Z$+jets processes 
are simulated using the {\alpgen} leading-order MC
generator~\cite{alpgen}, and those of diboson processes ($WW$, $WZ$
and $ZZ$) are modeled using {\pythia}~\cite{pythia}. The {\ttbar}
process is modeled using {\pythia} in the CDF analysis and by
{\alpgen} in the D0 analysis. Higgs-boson processes are modeled using
simulated events generated with {\pythia} for a Higgs-boson mass of
$m_H=125$~GeV.  The D0 analysis models the distributions and their shape
uncertainties of the $WH$ production process in the 2-jet, 2-$b$-tag
channel using simulated single-top-quark $t$-channel events
that have been shown to have the same distribution of the discriminant
output and to be only a small contamination to the $s$-channel signal. 
In all cases {\pythia} is used
to model proton remnants and simulate the
hadronization of all generated partons.  The mass of the top quark in
simulated events is set to
$m_t=172.5$~GeV, which is consistent with the
current Tevatron average value~\cite{topmassTeV}. All MC events are
processed through {\geant}-based detector simulations~\cite{geant}
and reconstructed by the same software packages used for the collider data.

Predictions for the normalization of simulated background-process
contributions are estimated using both simulation and data.
Data are used to normalize the $W$ plus
light-flavor and heavy-flavor jet contributions using enriched
$W$+jets data samples
that have negligible signal
content~\cite{d0_schannel,cdf-prd-2010,cdf_schannel_MET}.  All other
simulated background samples are normalized to the theoretical cross
sections at NLO combined with next-to-next-to-leading log (NNLL)
resummation~\cite{tchannel-kidonakis} for $t$-channel single-top-quark
production, at next-to-NLO~\cite{ttbar-xsec} for {\ttbar}, at
NLO~\cite{mcfm} for $Z$+jets and diboson production, and 
including all relevant higher-order QCD and electroweak corrections
for Higgs-boson production~\cite{higgs-xsec}.
Differences observed between simulated events and data in lepton and
jet reconstruction efficiencies, resolutions, jet-energy scale (JES), and
$b$-tagging efficiencies are adjusted in the simulation to match the
data, through correction functions obtained from measurements in
independent data samples.


We form multivariate discriminants,
optimized for separating the $s$-channel single-top-quark
signal events in each of the analysis samples from the larger
background contributions, to extract the cross section
measurements~\cite{s_vs_t_note}. The combined cross section
measurement is obtained using a Bayesian statistical 
analysis of the observed discriminant distributions from each sample,
comparing data to the modeled distributions for each of the
contributing signal and background processes~\cite{bayes-limits}.


A complete list of systematic uncertainties for the \ljets\ analyses is given in
Tab.~\ref{tab:tevsyst}. These can arise from uncertainties on
differential distributions (Dist) and their normalizations (Norm).
The CDF \MET+jets analysis has a similar set of systematic
uncertainties that are taken as fully correlated with the CDF \ljets analysis
except for the uncertainty related to the data-based background.
Sources of systematic uncertainty 
common to measurements of both collaborations are assumed to be 100\%
correlated, while other uncertainties are assumed to be 
uncorrelated.  The categories of uncertainty correspond generally to those
in Ref.~\cite{topmassTeV,ttbar_combi}, and can be summarized as follows:

\begin{description}

\item[Detector-specific luminosity uncertainty:] The component of the
  uncertainty on luminosity  that comes from the uncertainty on the
  acceptance and efficiency of the luminosity detector is taken
  as uncorrelated between the CDF~\cite{lumi-cdf} and
  D0~\cite{lumi-d0} measurements.

\item[Luminosity from cross section:] The portion of the uncertainty
  in luminosity that comes from uncertainties on the inelastic and
  diffractive cross sections is fully correlated between
  the CDF and D0 measurements. 

\item[Signal modeling:] The systematic uncertainty associated with
  uncertainties in the modeling of the single-top-quark signal,
  including uncertainties from the choice of the description of initial-
  and final-state QCD  
  radiation, and proton and antiproton parton density functions, 
  also covering uncertainties in the applied hadronization models,
  is taken as fully correlated between the CDF and D0 measurements.

\item[Background from simulation:] The systematic uncertainty associated
  with uncertainties in the modeling of various background
  contributions is taken as fully correlated between the CDF and D0
  measurements. This includes 
  uncertainties in \ttbar\ and diboson process normalizations
  originating from theoretical calculations. 

\item[Background based on data:] The systematic uncertainty associated
  with the modeling of various background sources
  obtained using data-driven methods is uncorrelated between the CDF
  and D0 measurements. This includes uncertainties on the
  normalization of $W$+jets, $Wb\bar{b}$, and $Wc\bar{c}$ events as well as
  uncertainties on the modeling of the 
  contributions and discriminant-variable shapes for the
  $W$+jets and QCD multijet production processes.  

\item[Detector modeling:] The systematic uncertainty on efficiencies
  for identifying reconstructed objects and to cover observed
  mismodeling of the data from the simulations is 
  uncorrelated between the CDF and D0 measurements.

\item[$b$-jet tagging:] The systematic uncertainty associated with the
  modeling of $b$-jet tagging efficiencies and 
  associated mistag rates is uncorrelated between
  the CDF~\cite{CDFbtag} and D0~\cite{D0btag} measurements.

\item[Jet energy scale (JES):] This systematic uncertainty originates from
   using calibration-data samples to establish the JES.
   For the CDF analyses, this corresponds to uncertainties associated with the
   $\eta$-dependent JES corrections, which are estimated using dijet
   events in data. For the D0 analysis, this includes uncertainties in calorimeter
   response for light jets, uncertainties from $\eta$- and
   $p_{T}$-dependent JES corrections, and other small
   contributions. This uncertainty is assumed to be uncorrelated
   between the CDF~\cite{cdf-jes} and 
   D0~\cite{d0-jes} measurements. 

\end{description}

\begin{table}[!h!btp]
\caption{
Systematic uncertainties associated with the CDF and the D0
single-top-quark $s$-channel cross section measurements in \ljets\
final states. The values
shown for each category indicate the range of uncertainties applied to
the predicted normalizations for signal and background contributions
over the full set of analysis samples from each experiment.  The black
dots indicate which categories contribute uncertainties on the shape
of the final multivariate discriminant output variable. It is also
noted if categories are treated as fully correlated between the two
experiments.} 
\begin{center}
\begin{tabular}{lccccc}\hline\hline
Systematic uncertainty          & \multicolumn{2}{c} {CDF} &  \multicolumn{2}{c} {D0} & Corre-\\
                                & Norm          & Dist         & Norm          & Dist & lated\\ \hline 
Lumi from detector        & 4.5\%         &               & 4.5\%         &      & No\\ 
Lumi from cross section\hspace{-0.05in}   & 4.0\%         &               & 4.0\%         &       & Yes \\ 
Signal modeling                 & 2--10\%   & $\bullet$     & 3--8\%   &       & Yes \\
Background (simulation)              &   2--12\%  & $\bullet$     & 2--11\%       &  $\bullet$     & Yes \\
Background (data)         & 15--40\%       & $\bullet$     & 19--50\%  & $\bullet$ & No\\
Detector modeling               & 2--10\%        & $\bullet$     & 1--5\%        &  $\bullet$     & No\\
$b$-jet-tagging                       & 10--30\%      &              & 5--40\%   & $\bullet$ &No \\ 
JES                            & 0--20\%        & $\bullet$     & 0--40\%   & $\bullet$ & No\\
\hline\hline
\end{tabular}
\end{center}
\label{tab:tevsyst}
\end{table}

The Bayesian posterior probability density as a function of
$s$-channel signal cross section($\sigma_{s}$) is given by 
\begin{equation}
p(\sigma_{s}) = \int L(\sigma_{s},\{\theta\}|{\rm data})\pi(\sigma_{s})
\Pi(\{ \theta \}) d\{\theta\} \, , \label{eq:posterior}
\end{equation}
where $L$ is the joint binned likelihood function for all channels
\begin{equation} 
\label{eq:LL}
L = \prod_{{\mathit i=\rm bins,\,channels}} \frac{(s_i + b_i)^{n_i}
e^{-(s_i+b_i)}}{n_i!}\,.
\end{equation}
The number of observed events in bin $i$ is $n_i$. 
$\{\theta\}$ is the set of nuisance parameters representing the
systematic uncertainties, and $\Pi(\{ \theta \})$ is the product of
the prior probability densities encoding the systematic uncertainties
on $\{ \theta \}$. The predictions for the number of signal events
$s_i$ and background events $b_i$ depend on the values of the
nuisance parameters that are integrated over in
Eq.~(\ref{eq:posterior}).  The  prior density for the signal cross
section, $\pi(\sigma_{s})$, is taken to be a uniform 
prior for non-negative cross sections.  We quote the measured cross
section as the value that maximizes its posterior likelihood, and the
uncertainty as the smallest interval that contains 68\% of the
integrated area of the posterior density.


Figure~\ref{fig:tevSigBkgSortInset.eps} shows the signal and
background expectations and the data as a function of $\log_{10} (s/b)$ of the
collected bins, for the combined CDF and D0 analyses.
%
\begin{figure}[!h!tbp]
\begin{center}
\includegraphics[width=0.48\textwidth]{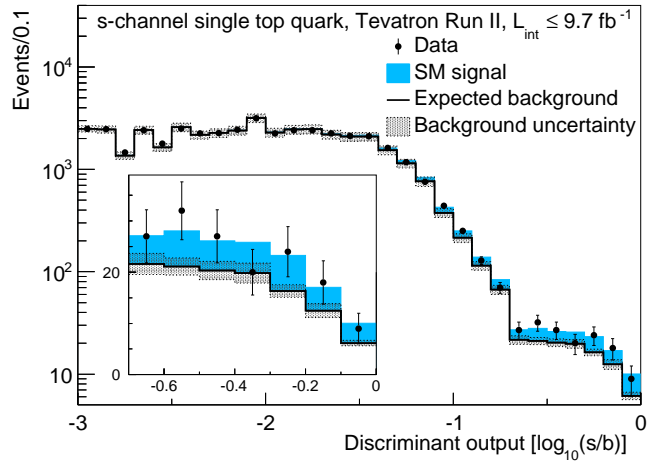}
\caption{(Color online) Distribution of the discriminant histograms,
  summed for bins with similar signal-to-background ratio ($s/b$). The
  expected sum of the backgrounds is shown by the unfilled histogram,
  and the uncertainty of the background is represented by the grey
  shaded band. The expected $s$-channel signal contribution is shown
  by a filled blue histogram.} 
\label{fig:tevSigBkgSortInset.eps}
\end{center}
\end{figure}
%
The extracted posterior probability
distribution for $\sigma_s$ is
presented in Fig.~\ref{fig:tevposterior}, and Fig.~\ref{fig:tevxsec}
gives a graphical presentation 
of the individual and combined measurements. 
%
\begin{figure}[!h!tbp]
\begin{center}
\includegraphics[width=0.48\textwidth]{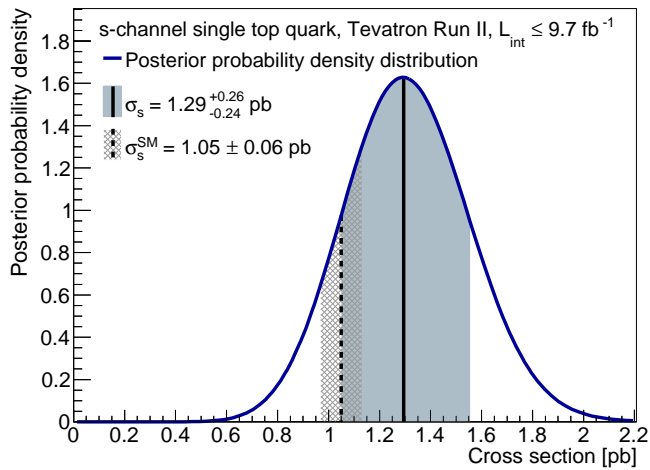}
\caption{The posterior probability distribution for the
  combination of the CDF and D0 analysis channels compared with the NLO+NNLL
  theoretical prediction~\cite{schannel-kidonakis}.}
\label{fig:tevposterior}
\end{center}
\end{figure}
%
%
\begin{figure}[!h!tbp]
\begin{center}
\includegraphics[width=0.48\textwidth]{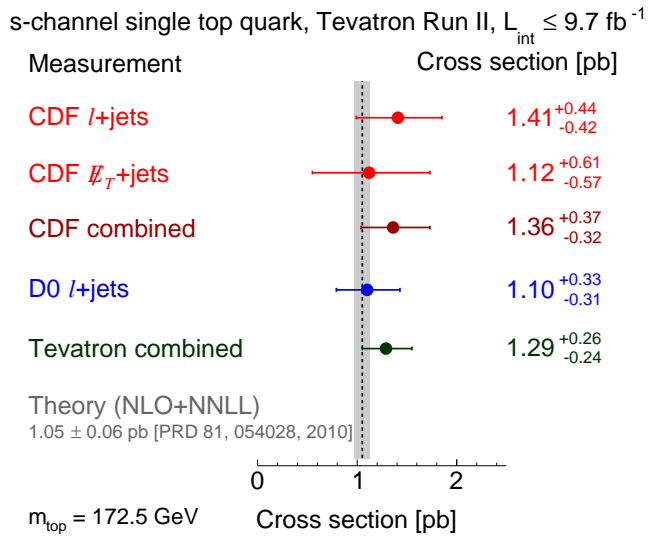}
\caption{(Color online) Measured single-top-quark $s$-channel
  production cross sections from each of the individual analyses and
  various combinations of these analyses compared with the NLO+NNLL
  theoretical prediction~\cite{schannel-kidonakis}. }
\label{fig:tevxsec}
\end{center}
\end{figure}
%
All measurements agree within their uncertainties with the SM
prediction, $\sigma_ s^{SM} = 1.05 \pm
0.06$~pb ($m_t=172.5$~GeV)~\cite{schannel-kidonakis}. The most
probable value for the combined  
cross section is $\sigma_s =
\xsectev^{\xsecteverrorup}_{\xsecteverrordown}$~pb for a top-quark
mass of 172.5~GeV.
The total expected uncertainty is 20\%, and the expected uncertainty
without considering systematic uncertainties is 14\%.  The dependence
of the measured value on the assumed value of the top-quark mass is
negligible compared to the uncertainty on the
measurement~\cite{cdf-prd-2010,d0_schannel}.  
 
The statistical significance of this result is quantified through a
calculated $p$-value based on an asymptotic log-likelihood ratio
approach (LLR)~\cite{loglhood}, including systematic
uncertainties. The $p$-value quantifies the 
probability that the measured value of the cross section or a larger
value could result from a background fluctuation in the absence of
signal.  The distributions of LLR resulting from
fits of simulated samples that include background-only, or
signal-plus-background, contributions are presented in
Fig.~\ref{fig:llr}. The probability to measure an $s$-channel cross
section of at least the observed value in the absence of signal is
$\probtev$, corresponding to a significance of $\sigmatev$ standard
deviations (s.\,d.), with a sensitivity expected from the SM of $\sigmaexp$
s.\,d. 

\begin{figure}[!h!tbp]
\begin{center}
\includegraphics[width=0.48\textwidth]{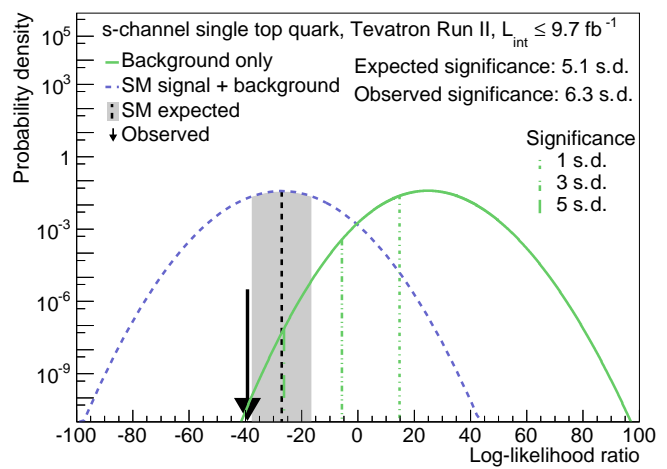}
\caption{(Color online) Log-likelihood ratios for the background-only
  (solid green line) and SM-signal-plus-background (dashed blue)
  hypotheses from the combined measurement. 
}
\label{fig:llr}
\end{center}
\end{figure}


In summary, we report the first observation of $s$-channel
single-top-quark production with a significance of \sigmatev~s.\,d. by
combining the CDF and D0 measurements. 
The combined value of the $s$-channel single-top-quark production
cross section 
is $\sigma_s = \xsectev^{\xsecteverrorup}_{\xsecteverrordown}$~pb, in agreement with
the SM expectation.


\section*{Acknowledgments}
We thank the Fermilab staff and technical staffs of the participating institutions for their vital contributions.
We acknowledge support from the
DOE and NSF (USA),
ARC (Australia),
CNPq, FAPERJ, FAPESP and FUNDUNESP (Brazil),
NSERC (Canada),
NSC, CAS and CNSF (China),
Colciencias (Colombia),
MSMT and GACR (Czech Republic),
the Academy of Finland,
CEA and CNRS/IN2P3 (France),
BMBF and DFG (Germany),
DAE and DST (India),
SFI (Ireland),
INFN (Italy),
MEXT (Japan),
the Korean World Class University Program and NRF (Korea),
CONACyT (Mexico),
FOM (Netherlands),
MON, NRC KI and RFBR (Russia),
the Slovak R\&D Agency,
the Ministerio de Ciencia e Innovaci\'{o}n, and Programa Consolider-Ingenio 2010 (Spain),
The Swedish Research Council (Sweden),
SNSF (Switzerland),
STFC and the Royal Society (United Kingdom),
the A.P. Sloan Foundation (USA),
and the EU community Marie Curie Fellowship contract 302103.
%




\begin{thebibliography}{99}
\bibitem{note}
The CDF and D0 experiments use a cylindrical coordinate system with origins in the
centers of the detectors, where $\theta$ and $\phi$ are the polar and
azimuthal angles, respectively, and pseudorapidity is 
$\eta= - \ln[\tan(\theta/2)]$. 
The transverse energy, as measured by the
calorimeter, is defined to be $E_T=E\sin\theta$. The missing $E_T$
(\METVEC) is defined by \METVEC$ = -\sum_iE_T^i{\hat n}_i, 
i={\rm calorimeter~tower~number}$, where ${\hat n}_i$ is a unit vector
perpendicular to the beam axis and pointing at the $i$th calorimeter
tower. ~\METVEC{} is corrected for high-energy muons and for
mismeasurements of jet energies. We define \MET=$|$\METVEC$|$. The
transverse momentum $p_T$ is defined to be $p\sin\theta$.
\bibitem{s_vs_t_note}
For the D0 collaboration this
is different from Ref.~\cite{d0_schannel}, where a combined $s$- and
$t$-channel discriminant was used to measure separately both single
top quark channels, without assuming their SM cross
sections when each is considered a background to the other process.
\end{thebibliography}
\end{document}

%
%
%